\documentclass{article}
\usepackage[no-math]{fontspec}
\usepackage[T1]{fontenc}
\usepackage{newunicodechar}
\usepackage{spconf,amsmath,graphicx,hyperref}
\usepackage{subcaption}
\usepackage{booktabs}
\usepackage{amssymb}
\usepackage{multirow}
\usepackage{xurl}
\usepackage{hyperref}
\usepackage{caption}
\usepackage{array,booktabs,makecell, multirow,tabularx}
\usepackage{float}
\usepackage{xcolor}
\usepackage{nicematrix}

\newfontfamily\arxivtokenfont{DejaVuSansMono.ttf}[Scale=MatchLowercase]
\newfontfamily\arxivromanfont{texgyretermes-regular.otf}
\newunicodechar{–}{\textendash}
\newunicodechar{ſ}{{\arxivtokenfont\char"017F}}
\newunicodechar{▁}{{\arxivtokenfont\char"2581}}
\newunicodechar{ș}{{\arxivromanfont\char"0219}}
\newcommand{\tokenLje}{{\arxivtokenfont\char"0409}}

\title{Analyzing Single-Bit Flips on LLMs and Their Backbone-Matched VLMs}
\name{
    Xiaomeng Wang$^{1}$,
    Martha Larson$^{1}$,
    Zhuoran Liu$^{2}$,
    Zhengyu Zhao$^{3}$
}

\address{
    $^{1}$Radboud University
    $^{2}$University of Amsterdam
    $^{3}$Xi'an Jiaotong University
}

\begin{document}
\maketitle
\begin{abstract}
Open-weight large language models (LLMs) enable offline weight analysis and are reused as backbones in vision-language models (VLMs), increasing the risk of bit-flip attacks that manipulate stored weight bits to induce malicious model behavior.
An adversary could identify vulnerable bits on a public LLM and later flip the same locations in a deployed LLM or its backbone-matched VLMs. 
A single-bit flip is particularly practical in this setting because it is easier to achieve under hardware fault attacks than multi-bit flips.
However, existing selectors are mostly limited to multi-bit flips, without in-depth analysis of single-bit flips.
In this paper, we adapt representative data-free and data-based bit selectors to identify candidate bits. 
We evaluate each bit independently on source LLMs and at the same logical coordinates in backbone-matched VLMs (without re-search), considering both performance degradation and prediction patterns across multiple inputs.
Results show impact of single-bit flips varies across selectors on LLMs and their backbone-matched VLMs.
A single-bit flip can make the model tend to produce a repeated output string across different inputs.
Our robustness analysis reveals vulnerabilities in terms of performance and prediction patterns.
\end{abstract}

\section{Introduction}
\label{sec:intro}
\begin{figure*}[t]
    \centering
    \includegraphics[width=\linewidth]{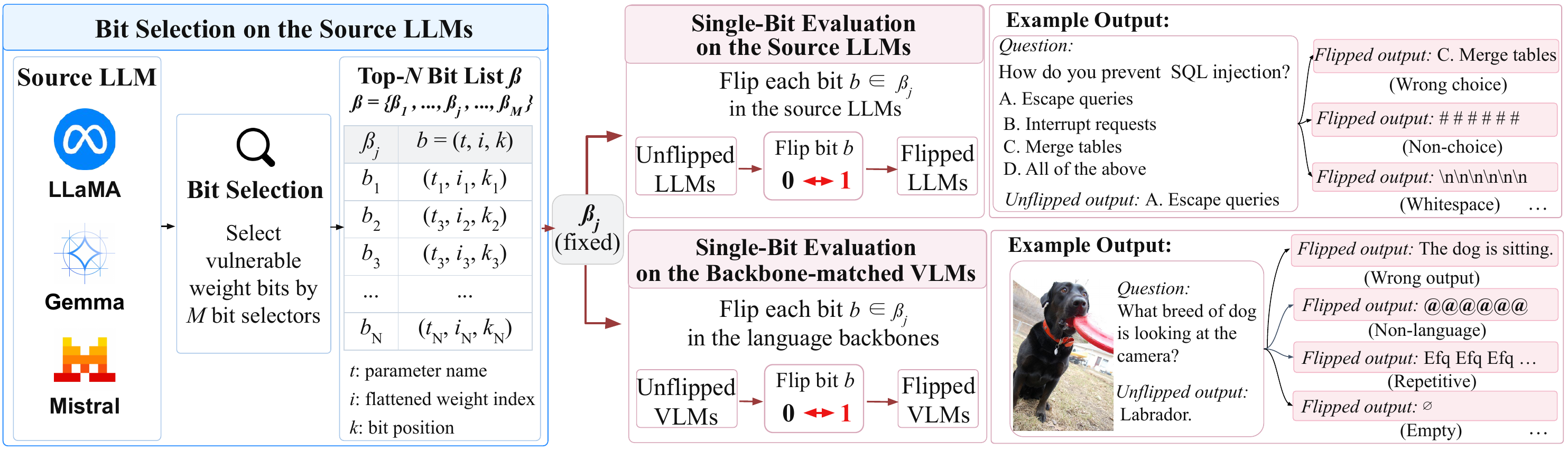}
    \caption{
Evaluation framework: bit selection and single-bit evaluation on the source LLMs and their backbone-matched VLMs, with one bit flipped at a time in evaluations.
\texttt{\textbackslash n} denotes newline;
$\varnothing$ denotes empty/whitespace strings excluding newlines.
}
    \label{fig:overview}
\end{figure*}

Bit-flip attacks (BFAs) are a form of fault injection that induce malicious behavior by flipping a small number of bits in model weights stored in main memory~\cite{rakin2019bfa,das2025attentionbreaker,almalky2025invisible,galil2026maximal}.
The increasing availability of open-weight large language models (LLMs) and their reuse as language backbones in vision-language models (VLMs) amplify the risk of bit-flip attacks across models.
This paper is motivated by the concern that an adversary may identify vulnerable locations on a public LLM and later modify the corresponding locations in the same LLM that is deployed or in a VLM that uses the LLM as a backbone. 
Single-bit flips are particularly realistic in this setting because manipulating a single vulnerable bit may be easier to realize under hardware fault attacks than attacks that require manipulation of multiple bits.
Single-bit attacks have been studied in DNNs under hardware fault settings~\cite{hong2019terminal}.

Previous works have left three gaps in our understanding of bit-flip attacks.
First, existing bit selectors are designed for attacks in which multiple selected bits are flipped together~\cite{rakin2019bfa,das2025attentionbreaker,almalky2025invisible,galil2026maximal}.
Consequently, they do not characterize the effects of a single bit in the bit set identified by existing selectors.
Prior work~\cite{guo2025sbfa} studies single-bit faults but focuses on the one bit that maximizes the damage of the attack.
Second, existing bit-flip studies focus on metrics that evaluate the outputs of the model.
The patterns of impact on the model outputs induced by a bit-flip attack across multiple inputs have not previously been studied.
These patterns might potentially either conceal or reveal the fact that the model has been attacked.
Third, prior bit-flip studies typically identify and manipulate vulnerable bits directly on the LLMs or VLMs.
This leaves open whether vulnerable locations identified offline on a public source LLM remain damaging when the same LLM architecture is reused as the language backbone of a VLM, without searching again on the VLM itself.

We fill these gaps by revisiting existing bit selectors and conducting experiments that make three contributions: 
(1) we evaluate the ability of individual bits identified by existing selectors to compromise LLM performance; 
(2) we go beyond model failure to also analyze the patterns of outputs that are produced across inputs; 
(3) we evaluate the impact of flipped LLMs on the VLMs that use them as backbones.

Our results show that accuracy drops vary across selectors and models, with some data-based selectors causing greater accuracy drops than the tested data-free ones.
Our analysis of impact patterns reveals that accuracy drop alone does not fully characterize the effects of bit-flip attacks. 
Some less damaging flips also produce frequent newline or whitespace predictions, suggesting an effectiveness–stealth trade-off. 
This risk also extends to backbone-matched VLMs: without re-search, flipping the same logical bit coordinates in the VLM language backbones, as selected on the LLMs, causes relative accuracy drops of at least 50\% in the evaluated configurations. 
Moreover, the flipped VLMs exhibit similar prediction patterns.

\section{Preliminaries}
\label{sec: preliminaries}
\noindent\textbf{Threat model.}
\emph{Attacker objective.}
We study untargeted attacks that degrade task accuracy without targeting specific outputs.
\emph{Attacker knowledge.}
The attacker has white-box access to the source LLM's architecture and weights and can perform forward/backward passes.
The attacker has gray-box access to the VLMs using the source LLM as backbone, where the attacker knows the target language backbone's identity and its logical parameters.
Target weight values, gradients, activations, and query feedback are unavailable for bit selection or refinement.
\emph{Attacker resources.}
In the data-free setting, the attacker has no real supporting data for bit selection but may generate random inputs~\cite{galil2026maximal}.
In the data-based setting, the attacker additionally has access to real text from a supporting corpus, such as a public dataset~\cite{rakin2019bfa,das2025attentionbreaker,almalky2025invisible}.
\emph{Attacker capabilities.}
The attacker can flip exactly one bit per attack in either the source LLM or the target VLM's backbone.
We evaluate logical faults; physical realization is out of scope.

\noindent\textbf{Weight quantization and bit flips.}
For the floating-point weight tensor $\mathbf{W}_l^{\mathrm{fp}}$ in Transformer layer $l$, we use symmetric per-tensor quantization to $N_q$-bit signed integers, where $N_q$ is the quantization bit width (e.g., 4, 8 for INT4, INT8).
We use $N_q=8$ in our experiments, with $Q_{\max}=2^{N_q-1}-1$, and scale $\Delta_l=\max|\mathbf{W}_l^{\mathrm{fp}}|/Q_{\max}$:
\begin{equation}
    \mathbf{q}_l
    = \operatorname{clip}\!\left(
        \operatorname{round}\!\left(
            \frac{\mathbf{W}_l^{\mathrm{fp}}}{\Delta_l}
        \right),
        -Q_{\max},Q_{\max}
    \right).
    \label{eq:quant_integer}
\end{equation}
Inference uses $\mathbf{W}_l^{\mathrm{q}}=\Delta_l\mathbf{q}_l$.
Each $q_{l,i}$ has two's-complement bits $b_k\in\{0,1\}$, with sign bit $b_{N_q-1}$:
\begin{equation}
    q_{l,i}
    = -2^{N_q-1}b_{N_q-1}
      + \sum_{k=0}^{N_q-2}2^k b_k.
    \label{eq:twos_complement}
\end{equation}
A fault flips exactly one bit and dequantizes the resulting integer using the original tensor scale.
We quantize source and target tensors independently with their own scales and transfer logical weight-bit coordinates, not physical memory addresses.
The same coordinate can thus induce different numerical perturbations in weight values across models due to differences in their quantized scales.

\section{Methodology for Single-Bit Analysis in LLMs and backbone-matched VLMs}
\label{sec:method}
Figure~\ref{fig:overview} shows our evaluation framework.
We first use bit selectors to generate and fix candidate bit lists for the source LLMs.
We then evaluate single-bit flips on both the source LLMs and their backbone-matched VLMs at the same logical coordinates, without re-search.

\noindent\textbf{Bit Selection on the Source LLM.}
We group selectors as described in the threat model in~\ref{sec: preliminaries}.
For a source LLM, each of the $M$ (here $M=3$) selectors $\mathcal{S}_j$ produces a candidate bit list
$\beta_j=(b_1,\ldots,b_N)$ using inputs $\mathcal{D}_j$,
where $\mathcal{D}_j$ is empty when no input is required for data-free selectors. Here $N=5$.
We denote the set of candidate lists as
$\beta=\{\beta_1,\ldots,\beta_j,\ldots,\beta_M\}$.
A logical coordinate $b_i=(t_i,i_i,k_i)$ specifies the canonical parameter name $t_i$, the flattened weight index $i_i$, and the bit position $k_i$.
The resulting lists $\beta_j$ are fixed during evaluation.

\noindent\textbf{Single-Bit Evaluation on LLMs and VLMs.}
We evaluate each bit $b\in\beta_j$ in the fixed candidate list on the source LLM and its backbone-matched VLMs.
The VLM language backbones preserve the source LLM's logical parameter organization but may differ in weight values.
We reuse the same coordinates $b=(t,i,k)$ in VLMs without re-search.
For each coordinate and model, we restore the original weights and apply exactly one bit flip using the fault operation defined in~\ref{sec: preliminaries}.
In VLMs, the flip is confined to the language backbone, leaving all other components unchanged.
Each evaluation measures a single fault rather than the cumulative effect of multiple flips.

\begin{table*}[!t]
\centering
\captionsetup{position=top,skip=0.5pt}
\caption{%
LLM and VLM performance under single-bit flips, measured by Perplexity on WikiText-2 and RAD on MMLU-Eval (LLMs) or VLM-Eval (VLMs).
Perplexity measures language prediction quality, and RAD measures the accuracy drop from the unflipped baseline; lower is better.
Columns show metric ranges, with entries counting top-64 bits in each range.
Counts are exact for LLMs and min--max across VLMs per backbone.
Wide Gemma-2 PPL ranges reflect baselines spanning 100.
}
\label{tab:llm_vlm_ppl_rad}
\label{tab:bit_flip_ppl_rad}
\label{tab:vlm_rad_summary}

\begingroup
\footnotesize
\setlength{\tabcolsep}{2pt}
\renewcommand{\arraystretch}{0.90}
\setlength{\aboverulesep}{0.15ex}
\setlength{\belowrulesep}{0.15ex}
\setlength{\dbltextfloatsep}{2pt}
\renewcommand{\arraystretch}{1.05}
\begin{tabular*}{\textwidth}{%
    @{\extracolsep{\fill}}
    l*{6}{c}@{\hspace{8pt}}*{6}{c}@{}
}
\toprule

\multirow{2}{*}{Selector}
& \multicolumn{3}{c}{Perplexity (Wikitext-2)}
& \multicolumn{3}{c}{RAD (MMLU-Eval)}
& \multicolumn{3}{c}{Perplexity (Wikitext-2)}
& \multicolumn{3}{c}{RAD (VLM-Eval)} \\
\cmidrule(lr){2-4}
\cmidrule(lr){5-7}
\cmidrule(lr){8-10}
\cmidrule(lr){11-13}

& $[1,100)$
& $[100,10^{3})$
& $[10^{3},+\infty)$
& $(-\infty,0.1)$
& $[0.1,0.5)$
& $[0.5,1]$
& $[1,100)$
& $[100,10^{3})$
& $[10^{3},+\infty)$
& $(-\infty,0.1)$
& $[0.1,0.5)$
& $[0.5,1]$ \\

\midrule
& \multicolumn{6}{l}{\textbf{Llama-2-7B}}
& \multicolumn{6}{l}{\textbf{Llama-2-7B-based VLMs}} \\

DNL
& 64 & 0 & 0 & 64 & 0 & 0
& 64--64 & 0--0 & 0--0 & 64--64 & 0--0 & 0--0 \\

1-DNL
& 56 & 3 & 5 & 52 & 12 & 0
& 55--55 & 2--2 & 7--7 & 52--55 & 0--3 & 9--9 \\

GDF
& 62 & 1 & 1 & 59 & 4 & 1
& 60--60 & 2--3 & 1--2 & 59--60 & 0--1 & 4--4 \\

AB
& 0 & 3 & 61 & 0 & 63 & 1
& 0--0 & 1--1 & 63--63 & 0--0 & 0--0 & 64--64 \\

PBS
& 62 & 0 & 2 & 60 & 4 & 0
& 62--62 & 0--0 & 2--2 & 62--62 & 0--0 & 2--2 \\

\midrule
& \multicolumn{6}{l}{\textbf{Gemma-2-2B}}
& \multicolumn{6}{l}{\textbf{Gemma-2-2B-based VLMs}} \\

DNL
& 64 & 0 & 0 & 64 & 0 & 0
& 0--64 & 0--64 & 0--0 & 64--64 & 0--0 & 0--0 \\

1-DNL
& 64 & 0 & 0 & 64 & 0 & 0
& 0--62 & 2--64 & 0--0 & 63--64 & 0--1 & 0--0 \\

GDF
& 64 & 0 & 0 & 64 & 0 & 0
& 0--64 & 0--64 & 0--0 & 64--64 & 0--0 & 0--0 \\

AB
& 64 & 0 & 0 & 64 & 0 & 0
& 0--57 & 7--64 & 0--0 & 64--64 & 0--0 & 0--0 \\

PBS
& 60 & 1 & 3 & 59 & 2 & 3
& 0--36 & 23--58 & 5--6 & 53--59 & 0--5 & 5--6 \\

\midrule
& \multicolumn{6}{l}{\textbf{Mistral-7B-v0.1}}
& \multicolumn{6}{l}{\textbf{Mistral-7B-based VLMs}} \\

DNL
& 63 & 0 & 1 & 63 & 0 & 1
& 63--63 & 0--0 & 1--1 & 63--63 & 0--0 & 1--1 \\

1-DNL
& 57 & 5 & 2 & 51 & 1 & 12
& 52--57 & 5--7 & 2--6 & 45--52 & 1--7 & 9--12 \\

GDF
& 63 & 0 & 1 & 63 & 0 & 1
& 63--63 & 0--0 & 1--1 & 62--63 & 0--1 & 1--1 \\

AB
& 2 & 14 & 48 & 1 & 1 & 62
& 1--2 & 2--9 & 54--60 & 0--2 & 0--1 & 62--63 \\

PBS
& 27 & 31 & 6 & 19 & 1 & 44
& 20--27 & 26--27 & 10--17 & 18--25 & 0--10 & 33--46 \\

\bottomrule
\end{tabular*}
\par
\endgroup
\end{table*}
\vspace{-4pt}

\begin{table*}[]
\centering
\renewcommand{\thetable}{3}
\captionsetup{position=top,skip=0.5pt}
\caption{%
Output string dominance ($D_b$) and most common strings for the top-64 bits per model--selector pair, computed on MMLU-Eval for LLMs and VLM-Eval for VLMs.
$D_b$ is the percentage of inputs predicting the most frequent output string; higher values mean more inputs predict the same output string.
Columns show $D_b$ bins, and entries count bits per bin; counts are for LLMs and min--max across nine VLM--dataset pairs per backbone.
Most common strings are shown for $D_b>0.5$.
\texttt{\textbackslash n} denotes newline, $\varnothing$ empty/whitespace, and \texttt{</s>} / \texttt{<pad>} EOS / padding tokens.
``Other'' denotes remaining distinct strings.
}
\label{tab:llm_vlm_token_patterns}
\label{tab:token_attractor_patterns}
\label{tab:vlm_token_dominance_distributions}

\begingroup
\scriptsize
\setlength{\tabcolsep}{1.3pt}
\renewcommand{\arraystretch}{1.00}
\setlength{\aboverulesep}{0.15ex}
\setlength{\belowrulesep}{0.15ex}

\resizebox{\textwidth}{!}{%
\begin{tabular}{@{}lccc l @{\hspace{7pt}} ccc l@{}}
\toprule
\multirow{2}{*}{Selector}
& \multicolumn{3}{c}{Output string dominance}
& \multirow{2}{*}{Most common strings (\# bits)}
& \multicolumn{3}{c}{Output string dominance}
& \multirow{2}{*}{Most common strings (min--max \# bits)} \\
\cmidrule(lr){2-4}
\cmidrule(lr){6-8}
& $[0,0.5]$ & $(0.5,0.9)$ & $[0.9,1]$
&
& $[0,0.5]$ & $(0.5,0.9)$ & $[0.9,1]$
& \\
\midrule

& \multicolumn{4}{l}{\textbf{Llama-2-7B}}
& \multicolumn{4}{l}{\textbf{Llama-2-7B-based VLMs}} \\

DNL
& 64 & 0 & 0 & ---
& 64--64 & 0--0 & 0--0 & --- \\

1-DNL
& 58 & 2 & 4
& \texttt{\textbackslash n} (4), \texttt{(} (1), \texttt{s} (1)
& 55--58 & 2--4 & 4--6
& \texttt{</s>} (2--3), \texttt{-} (1--1),
  \texttt{\textbackslash n} (0--2) \\

GDF
& 64 & 0 & 0 & ---
& 60--62 & 0--3 & 0--2
& \texttt{The} (1--2), \texttt{what} (0--1),
  $\varnothing$ (0--1) \\

AB
& 11 & 28 & 25
& \texttt{\textbackslash n} (14), $\varnothing$ (9),
  \texttt{a} (2); Other 24 (28)
& 2--6 & 19--28 & 34--42
& $\varnothing$ (7--9), \texttt{</s>} (5--8),
  \texttt{The} (5--8) \\

PBS
& 46 & 17 & 1
& \texttt{A} (16), \texttt{,} (1), \tokenLje (1)
& 62--62 & 0--2 & 0--2
& \texttt{,} (0--1), $\varnothing$ (0--1),
  \tokenLje (0--1) \\

\midrule

& \multicolumn{4}{l}{\textbf{Gemma-2-2B}}
& \multicolumn{4}{l}{\textbf{Gemma-2-2B-based VLMs}} \\

DNL
& 64 & 0 & 0 & ---
& 64--64 & 0--0 & 0--0 & --- \\

1-DNL
& 64 & 0 & 0 & ---
& 64--64 & 0--0 & 0--0 & --- \\

GDF
& 64 & 0 & 0 & ---
& 64--64 & 0--0 & 0--0 & --- \\

AB
& 64 & 0 & 0 & ---
& 64--64 & 0--0 & 0--0 & --- \\

PBS
& 59 & 3 & 2
& \texttt{<pad>} (2), \texttt{A} (1),
  \texttt{B} (1), \texttt{Theſe} (1)
& 59--61 & 0--3 & 2--4
& \texttt{<pad>} (2--2), \texttt{▁Theſe} (1--1),
  \texttt{▁ſeveral} (0--1) \\

\midrule

& \multicolumn{4}{l}{\textbf{Mistral-7B-v0.1}}
& \multicolumn{4}{l}{\textbf{Mistral-7B-based VLMs}} \\

DNL
& 63 & 1 & 0
& $\varnothing$ (1)
& 63--64 & 0--1 & 0--1
& \texttt{\textbackslash n} (0--1) \\

1-DNL
& 55 & 9 & 0
& \texttt{D} (6), \texttt{\textbackslash n} (1),
  \texttt{The} (1), $\varnothing$ (1)
& 56--61 & 1--7 & 1--3
& \texttt{The} (0--6), \texttt{\textbackslash n} (0--2),
  \texttt{to} (0--1) \\

GDF
& 63 & 1 & 0
& $\varnothing$ (1)
& 63--64 & 0--1 & 0--1
& \texttt{\textbackslash n} (0--1) \\

AB
& 32 & 20 & 12
& \texttt{\textbackslash n} (17), $\varnothing$ (4),
  \texttt{A} (3); Other 8 (8)
& 4--11 & 19--35 & 18--40
& \texttt{\textbackslash n} (5--15), $\varnothing$ (7--15),
  \texttt{</s>} (1--14) \\

PBS
& 51 & 11 & 2
& \texttt{\textbackslash n} (6), \texttt{D} (5),
  \texttt{B} (1), \texttt{bekan} (1)
& 28--43 & 10--21 & 8--15
& \texttt{The} (2--12), \texttt{\textbackslash n} (2--10),
  $\varnothing$ (1--4) \\

\bottomrule
\end{tabular}%
}
\endgroup
\end{table*}
\vspace{-5pt}

\section{Experiments}
\label{sec:experiments}
\subsection{Experimental Setup}
\label{sec:single-bit-setup}
\noindent\textbf{Models.}
Source LLMs are meta-llama/Llama-2-7b-hf
\cite{touvron2023llama2}, Google/gemma2-2b
\cite{team2024gemma}, and mistralai/Mistral-7B-v0.1~\cite{jiang2023Mistral7}.
Llama-2-7B is paired with three Prismatic VLMs
\cite{karamcheti2024prismatic}: prism-clip+7b, prism-siglip+7b, and prism-dinosiglip+7b.
Gemma-2-2B is paired with three PaliGemma checkpoints
\cite{beyer2024paligemma}: google/paligemma-3b-pt-224, google/paligemma-3b-mix-224, and google/paligemma-3b-ft-textvqa-448.
Mistral-7B-v0.1 is paired with Idefics2-8B-Base, Idefics2-8B\cite{laurenccon2024matters}, and BakLLaVA-1~\cite{skunkworksai2023bakllava}.
Tensors are matched by canonical parameter name and shape, preserving flattened weight indices and bit positions despite possible differences in weight values.
Inference uses Bfloat16 (BF16).
Before each single-bit evaluation, we restore the original weights, quantize the attacked tensor to signed INT8, flip the selected bit, and dequantize it to BF16, leaving other tensors unchanged.
The quantization-matched unflipped baseline follows the same procedure without the flip.
Token embeddings and language-model heads are excluded from bit selection.

\noindent\textbf{Bit selectors.}
We retain the top 64 candidate bits per LLM from each selector: data-free DNL and 1-DNL~\cite{galil2026maximal}, and data-based GDF~\cite{almalky2025invisible}, AB~\cite{das2025attentionbreaker}, and PBS~\cite{rakin2019bfa}.
Data-based selectors use 16 WikiText-2 test records~\cite{merity2017wikitext}, filtered to at least 25 words and truncated to 384 tokens.
DNL/1-DNL search attention and MLP projections in the first five decoder blocks; GDF uses \(k_{\mathrm{top}}=8\) with a non-final block selected by layer vulnerability; PBS uses \(k_{\mathrm{top}}=10\) for at most 64 iterations.

\noindent\textbf{Datasets.}
For LLM evaluation, we use 128 disjoint WikiText-2 records and MMLU-Eval, with 20 test inputs from each of 57 MMLU~\cite{mmlu} subjects (1,140 total) and five-shot demonstrations from the corresponding development splits.
For VLM evaluation, we use VLM-Eval, with 100 sampled image--question inputs from each of TextVQA~\cite{singh2019textvqa}, VQAv2~\cite{goyal2017vqav2}, and GQA~\cite{hudson2019gqa} validations (300 total), using the prompt ``Answer the question using a single word or phrase.'' and greedy decoding with at most 16 tokens.
MMLU-Eval has multiple-choice questions with four predefined answer options, whereas VLM-Eval has open-ended visual questions.

\noindent\textbf{Metrics.}
We measure performance degradation using perplexity~\cite{hu2024can} on WikiText-2 evaluation set and Relative Accuracy Drop (RAD): $\mathrm{RAD} = 1 -\frac{\mathrm{Acc}_{\text{flipped}}}{\mathrm{Acc}_{\text{unflipped}}},$
where both accuracies use the same tensor quantization and dequantization.
We use micro-averaged accuracy for MMLU, official normalized soft VQA accuracy for TextVQA and VQAv2, and normalized exact-match accuracy for GQA.
We record the top-1 token from the full-vocabulary logits at the first answer position for each input and count predictions by exact token ID.
We define output string dominance $D_b$ as the proportion of inputs producing the output string that is most frequent produced across all inputs.
$D_b>0.5$ means that more than half of the inputs predict the same output string.

\begin{table}[H]
\centering
\renewcommand{\thetable}{2}
\captionsetup{skip=0.5pt}
\caption{Unflipped baselines for LLMs and VLMs.
Perplexity is evaluated on WikiText-2.
LLM accuracy is evaluated on MMLU-Eval, while VLM accuracy is evaluated on TextVQA, VQAv2, and GQA (collectively referred to as VLM-Eval).
All accuracies are in percentage.}
\label{tab:unflipped_baselines}

\footnotesize
\setlength{\tabcolsep}{1.8pt}
\renewcommand{\arraystretch}{1}
\resizebox{\columnwidth}{!}{%
\begin{tabular}{@{}llccccc@{}}
\toprule
Backbone & Model & Perplexity & MMLU-Eval & TextVQA & VQAv2 & GQA \\
\midrule

\multicolumn{7}{@{}l}{\textbf{LLMs}} \\
-- & Llama-2-7B
& 8.43 & 47.1 & -- & -- & -- \\
-- & Gemma-2-2B
& 13.58 & 54.4 & -- & -- & -- \\
-- & Mistral-7B
& 8.95 & 64.1 & -- & -- & -- \\

\midrule
\multicolumn{7}{@{}l}{\textbf{VLMs}} \\

\multirow{3}{*}{Llama-2}
& prism-clip+7b
& 13.37 & -- & 54.2 & 80.3 & 69.0 \\
& prism-siglip+7b
& 13.02 & -- & 58.3 & 80.3 & 71.0 \\
& prism-dinosiglip+7b
& 13.28 & -- & 54.5 & 83.4 & 76.0 \\

\multirow{3}{*}{Gemma-2}
& PaliGemma2-3B-PT-224
& 178.24 & -- & 8.4 & 42.1 & 28.0 \\
& PaliGemma2-3B-Mix-224
& 96.96 & -- & 51.6 & 80.2 & 56.0 \\
& PaliGemma2-3B-Mix-448
& 102.91 & -- & 63.8 & 82.3 & 60.0 \\

\multirow{3}{*}{Mistral}
& Idefics2-8B-Base
& 8.88 & -- & 32.0 & 36.2 & 21.0 \\
& Idefics2-8B
& 9.75 & -- & 67.1 & 72.3 & 51.0 \\
& BakLLaVA-1
& 28.67 & -- & 46.8 & 72.3 & 67.0 \\

\bottomrule
\end{tabular}
}
\end{table}

\subsection{Experimental Results}
\noindent\textbf{Varying degradation under single-bit flips.}
Table~\ref{tab:bit_flip_ppl_rad} reports performance degradation in source LLMs and backbone-matched VLMs using the same fixed bit lists, where the uplipped accuracies used in RADs are shown in Table~\ref{tab:unflipped_baselines}.
On Llama-2-7B, 63/64 AB-selected bits yield $\mathrm{RAD}\in[0.1,0.5)$ and only 1/64 yields $\mathrm{RAD}\geq0.5$, although 61/64 yield $\mathrm{PPL}\geq10^3$.
In the backbone-matched VLMs, all 64 AB-selected bits yield $\mathrm{RAD}\geq0.5$ over the evaluated configurations.
On Mistral-7B-v0.1, 62/64 AB-selected bits reach this RAD threshold, compared with 62--63/64 in its VLMs.
For Gemma-2-based VLMs, only PBS reaches this threshold (5--6/64 bits).
These results show that degradation varies across selectors and models, and single-bit flips identified on source LLMs remain effective on backbone-matched VLMs

\noindent\textbf{Prediction patterns beyond degradation.}
Table~\ref{tab:token_attractor_patterns} shows prediction patterns on LLMs (MMLU-Eval) and VLMs (VLM-Eval) using output string dominance and most common strings.
For PBS, at least 14/64 bits on Llama-2-7B have $D_b>0.5$ with $\mathrm{RAD}<0.1$, whereas at least 31/64 bits on Mistral-7B-v0.1 have $\mathrm{RAD}\geq0.5$ with $D_b\leq0.5$.
A small accuracy drop can occur with the same output strings across inputs, while a large drop may not.
For LLMs, we also measure the non-choice rate, i.e., the percentage of output string outside A--D.
For PBS, 61/64 bits on Llama-2-7B have zero non-choice rate, and 16/18 bits with $D_b>0.5$ most commonly predict \texttt{A}.
In contrast, AB yields at least 90\% non-choice predictions for 60/64 bits on Llama-2-7B and 29/64 on Mistral-7B-v0.1.
Predictions outside A--D may involve different tokens rather than the same token.
In the VLM-Eval, AB yields $D_b>0.5$ for 58--62/64 bits in Llama-2-based VLMs and 53--60/64 in Mistral-based VLMs, often dominated by newline or \texttt{</s>}.
Overall, the identity of output string predictions across inputs helps reveal single-bit effects beyond accuracy drop.

\section{Conclusion and Outlook}
\label{sec:conclusion}
We evaluate single-bit flips on LLMs and their backbone-matched VLMs using data-free and data-based selectors, examining both performance degradation and prediction patterns. 
The results show that bits selected on source LLMs can also degrade their backbone-matched VLMs at the same logical coordinates without re-search. 
Performance degradation varies across selectors on LLMs and their backbone-matched VLMs. 
Beyond performance degradation, prediction pattern analysis shows that some single-bit flips cause one output string to dominate across different inputs, revealing patterns not captured by accuracy alone.

These results also provide implications for improving model robustness. 
Across our evaluations, Gemma-2-2B and its backbone-matched VLMs are less vulnerable overall than the other tested models, although PBS still identifies damaging bits. 
We hypothesize that differences in activation outliers partly explain this variation, motivated by prior findings on massive activations~\cite{sun2024massive} and the use of activation outliers to identify vulnerable weights~\cite{almalky2025invisible}. 
Future work could test this hypothesis and investigate whether such properties can guide bit-flip-resistant model design or selective protection of vulnerable parameters.




\bibliographystyle{IEEEbib}
\bibliography{strings,reference}

\end{document}